# Reciprocally induced coevolution:
# A computational metaphor in Mathematics

**SIBY ABRAHAM**

Dept of Maths & Stats,

G. N. Khalsa College,

University of Mumbai,

Mumbai, India.

sibyam@gmail.com

**SUGATA SANYAL**

School of Tech & Comp Sc.,

Tata Inst. of Fundamental Research,

Mumbai, India.

sanyals@gmail.com

**MUKUND SANGLIKAR**

Dept of Mathematics,

Mithibai College,

University of Mumbai,

Mumbai, India.

masanglikar@rediffmail.com

**Abstract**
Natural phenomenon of coevolution is the reciprocally induced evolutionary change between two or more species or population. Though this biological occurrence is a natural fact, there are only few attempts to use this as a simile in computation. This paper is an attempt to introduce reciprocally induced coevolution as a mechanism to counter problems faced by a typical genetic algorithm applied as an optimization technique. The domain selected for testing the efficacy of the procedure is the process of finding numerical solutions of Diophantine equations. Diophantine equations are polynomial equations in Mathematics where only integer solutions are sought. Such equations and its solutions are significant in three aspects-(i) historically they are important as Hilbert's tenth problem with a background of more than twenty six centuries ;(ii) there are many modern application areas of Diophantine equations like public key cryptography and data dependency in super computers (iii) it has been proved that there does not exist any general method to find solutions of such equations. The proposed procedure has been tested with Diophantine equations with different powers and different number of variables.

## 1.0 Introduction

Coevolution is defined, in evolutionary biology, as reciprocally induced evolutionary change between two or more species or populations [Price 1998]. It is the mutual influence exerted by species on one another in the evolutionary process [Carter 2005]. Coevolution can be broadly divided into three types: predator prey coevolution, mutualism and host parasite coevolution [Simms 1996].

In predator-prey relationships, the predators, who have to eat prey for survival, are often acquired prey specific characteristics like acute senses, claws, teeth, fangs, stingers or poison that help them catch and chew the prey. Preys, at the same time, are endowed with defense mechanism like mobbing, alarm calls, poison or an ability to mimic another species [Hochberg & Van Baalen]. Predator and prey both evolve together and each one is part of the others environment. Predator dies if it does not get food and so does whatever possible to get the prey. At the same time, prey will be eaten by the predator and so does whatever is needed to escape from the predators. Some of the common predator-prey duos in nature are lion and zebra, bear and fish, fox and rabbit, bear and berry, rabbit and lettuce, and grasshopper and leaf [Necsi 2010].

In mutualism, two organisms interact on a regular basis and both of them acquire a better chance of survival because of this co-habitat. Such phenomena are very much visible in nature. One such instance is the special relationship that exists between insect-pollinated plants and their insect pollinators, which is exemplified in the case of acacia plants and acacia ants. Acacia plants have large and hollow thorns and acacia ants live in those thorns. These ants are survived by eating a substance produced by the plants at the tip of its leaflets. The ants protect the plants from attacking herbivores and outside seedlings. Such 'service' and 'reward' [Heil et al2010] reciprocity in mutualism helps both species, as they are dependent on one another for survival.

Host parasite coevolution is defined as the reciprocal genetic change of two species by selective reciprocal influences. Nature offers a number of examples in which organisms evolve defenses to attacks of parasites. These parasites in turn, find ways to circumvent these defenses and attack the organisms with greater vigor. This results in the evolution of new defenses by the host. This ever-rising strategy of continued, fresh attacks by parasites and formation of new defensive strategy by host evolve in the process. This 'biological arms race' [Kniskern & Rausher 2001] [Hillis 1992] of fight of dominance by parasites and host continues in nature. Host parasite coevolution is considered to be one of the main reasons of earth's diversity. Medically relevant diseases like malaria, AIDS and Influenza are caused by coevolving parasites.

Though the concept of host parasite coevolution as a natural phenomenon has been prevalent, Hillis [Hillis] was the one, who introduced this biological concept into research domain of computer science. Coevolution is better suited to approach problems with large Cartesian-product spaces because of the inherent 'arms race' offered by it drives the search to untried and untested parts of the search space [Wiegand 2001].

There have been a number of attempts to use coevolution to solve problems in different domains. Hillis [Hillis 1992] used host parasite coevolution methodology to find a solution to Sorting Networks [Knuth 1997] by taking a population of sorting networks as hosts and a population of networks coevolved on the same grid as a population of parasites. Stanley and Miikkulainen [Stanley & Miikkulainen 2002] applied coevolution in a complex competitive robot duel domain and introduced a procedure called dominance tournament, which showed how different coevolution run continued to innovate for different periods of time and identified the best individuals found during the runs. Tsujimora [Tsujimora 2001] proposed a coevolution process for Job Shop Scheduling by offering a unique fitness functions to the operational genetic algorithm. Bui et al [Bui et al 2005] described a procedure to solve a special class of rotated problems, which coevolved the mapping represented by a population of matrices in parallel with the genotypes. Dejong [Dejong 2002] described the development of representation of a Genetic Algorithm as a coevolution problem by coevolving building blocks and assemblies. Potter and Dejong [Potter & Dejong 2000] employed coevolution to train dynamically structured recurrent neural networks. Game theory is one of the areas, where coevolution has found many applications in recent times. Ficci [Ficci 2005] used evolutionary game theory to investigate the dynamics and equilibrium of selection methods in coevolutionary algorithms and demonstrate that only Boltzmann selection converged onto polymorphic Nash equilibrium. Ficci also [Ficci 1998] applied coevolution to reformulate pursuer-evader game as one dimensional time series prediction game to capture fundamental aspects of communication. Viswanthan [Viswanathan 2007] proposed a coevolution based design strategy for representation of a Genetic Algorithm and showed that Genotype-Phenotype map that is biasfree is formally equivalent to Nash equilibrium in a non-cooperative multiplayer game. Juille [Juille 1998] introduced the concept of coevolutionary learning as a search procedure involving a population of learners coevolving with a population of problems such that continuous progress resulted from this interaction. Pagie and Mitchell [Pagie & Mitchell 2002] compared evolutionary and coevolutionary search for solutions to the density classification task for cellular automata and exhibited the superior nature of coevolutionary search.

This paper is an attempt to validate the effectiveness of host parasite coevolution in the realm of Mathematics. The case study used to bring home the point is the process of finding numerical solutions of a system of equations known as Diophantine equations using a genetic algorithm [Mitchell 1998]. The genetic algorithm uses genetic operators [Michalewich 1992] - selection, inversion and cross over- to generate new offspring [Goldberg 2006]. The procedure adopts a composite evaluation function to check the suitability of an individual in the population. The repeated values of the evaluation function are taken as attacks by parasites and the values of evaluation function of the new chromosomes generated are taken as the defenses of the host. The paper explains how the procedure of coevolution of attacks by parasites and defenses by host induce an evolutionary pressure to create better chromosomes and eventually resulting in finding the solution of a given Diophantine equation.

## 2. 0 Diophantine Equations
A Diophantine equation [Rosen 2000] is a polynomial equation, given by
$$f(a_1, a_2, ....., a_n, x_1, x_2, ......, x_n) = N \quad .........(1)$$
where $a_i$ and N are integers. The equation (1), which is christened after the third century Alexandria Mathematician Diophantus, may have:

- No nontrivial, integral solution as in the case of $x_1^2 + x_2^2 + x_3^2 = 0$ ...... (2)
- Finitely many integral solutions as in the case of $x_1^4 + x_2^4 - x_3^4 = 0$ ....... (3)
- Infinitely many integral solutions as in the case of $x_1^n + x_2^n - x_3^n = 0$ ..... (4)

Mathematicians over many centuries have shown interest in general Diophantine equations and its particular types and have attempted to find its solutions. For example, linear equation $a_1 * x_1 + a_2 * x_2 + ....... + a_n * x_n = N$ has solutions if and only if the greatest common divisor $(a_1, a_2, \ldots, a_n)$ divides N. The equation $x_1^n + x_2^n - x_3^n = 0$ is the equation in Fermat's last theorem, which he conjectured, had no integral, non-trivial solutions for n>2 [Rosen]. For n = 2, the equation reduces to $x_1^2 + x_2^2 - x_3^2 = 0$ whose solutions are known as Pythagorean triplets and are primitive in nature. There is at least one integral solution to the equation $x_1^2 + x_2^2 + x_3^2 + x_4^2 = N$, which Warring conjectured that, could be generalized to higher powers with 9 cubes, 19 fourth powers etc [Niven & Zuckerman 1972].

The individual attempts to demonstrate the importance of particular types of Diophantine equations and its solution were collectively posed by Hilbert in his famous tenth problem- whether there exists an algorithm to find a solution for the general Diophantine equations with integral coefficients? Decades later [Matiyasevich 19975] conclusively proved that it is impossible to obtain a general solution for (1). As the scope and importance of Diophantine equations are not restricted to the abstract realm of number theory but has applications in fields like Public key cryptosystems [Lin 1995] [Laih 1997], computable economics [Velu 2000] and theoretical computer Science [Ibarra 2004] [Guarari 1998], there have been attempts to solve (1) numerically.

This turned out to be a hard problem as the search space consists of $N^n$ elements resulting in attempts to find numerical solutions using Artificial Intelligence Techniques [Rich & Knight 1991] [Russell & Norvig 2003]. Literature mentions these attempts in [Abraham & Sanglikar 2001] [Abraham & Sanglikar 2007 a] [Abraham & Sanglikar 2007 b] [Abraham & Sanglikar 2008] [Abraham & Sanglikar 2009] [Abraham et al 2010] and [Abraham et al 2011].

### 3.0 Methodology used
The methodology described in the paper uses some keywords. We define those keywords as below:

**3.1 Gene:** Gene is an integer from the set {a+1, a+2, a+3,......, b} where a and b are supplied end values for the range of solution for a given Diophantine equation.

**3.2 Chromosome:** Chromosome is an r-tuple $(s_1, s_2, s_3, \ldots, s_r)$ where each element is a gene and r = b-a. For example, (1, 2, 3, .........., 25) is a chromosome. We label the set of all chromosomes by G and at any instant of evolutionary time; the set of chromosomes present at that time constitutes the population.

**3.3 Evaluation Function:** Evaluation Function is a mathematical function, which is automatically defined at every iteration. The selection or rejection of a candidate is based on the value of the evaluation function. We start with an evaluation function, which is given by

$$\text{eval} = \text{Abs}(N - (a_1 * x_1^{p_1} + a_2 * x_2^{p_2} + \ldots + a_n * x_n^{p_n})) \quad \ldots\ldots (5)$$

For example, for the equation $x_1^2 + x_2^2 + x_3^2 + x_4^2 + x_5^2 + x_6^2 = 544$, the initial evaluation function value of a chromosome (1, 2, 3, 4, 5, 6, 7, 8, 9, 10, 11, 12, 13, 14, 15, 16, 17, 18, 19, 20) is given by eval = $544 - (1^2 + 2^2 + 3^2 + 4^2 + 5^2 + 6^2)$.

**3.4 Inversion:** Inversion is an operator $M_{ij} : G \rightarrow G$. It is a unary operator and generates a new chromosome by swapping randomly selected genes within the parent chromosome. For example, if (1, 2, 3, *4,* 5, 6, 7, 8, 9, 10, 11, 12, 13, *14*, 15, 16, 17, 18, 19, 20) is a chromosome, then exchanging the gene **4** and gene **14**, we get a new chromosome (1, 2, 3, *14,* 5, 6, 7, 8, 9, 10, 11, 12, 13, *4*, 15, 16, 17, 18, 19, 20)

**3.5 Crossover:** Crossover is a binary operator C: G * G → G to generate new chromosomes. It applies on two chromosomes by generating a random point of crossover and swapping genes from thereon between the selected chromosomes to produce new chromosomes. For example, chromosomes (*1, 2, 3, 4, 5, 6, 7,* 8, 9, 10, 11, 12, 13, 14, 15, 16, 17, 18, 19, 20) and (*20, 19, 18, 17, 16, 15, 14*, 13, 12, 11, 10, 9, 8, 7, 6, 5, 4, 3, 2, 1) will on crossover generate two offsprings-(*1, 2, 3, 4, 5, 6, 7*, 13, 12, 11, 10, 9, 8, 7, 6, 5, 4, 3, 2, 1) and (*20, 19, 18, 17, 16, 15, 14*, 8, 9, 10, 11, 12, 13, 14, 15, 16, 17, 18, 19, 20) where the crossover point is '7'.

**3.6 Selection:** Selection is a genetic operator applied to choose a limited number of chromosomes among a collection of chromosomes based on their fitness values. Selected chromosomes constitute the new population to be part of the evolutionary process in the next generation.

**3.7 Procedure**: We start the procedure with a population of five chromosomes. These chromosomes are generated randomly based on the values 'a' and 'b' supplied as the range within which a solution is to be found. Initial evaluation function values of these chromosomes are calculated using the equation (5). Then, these 'eval' values are sorted. If there are any chromosome with 'eval' = 0, then that chromosome is taken as a solution for the given equation. Generating a random number between 0 and 1 incorporates randomness. If it is greater than the entered value of probability of inversion, then inversion is selected as the genetic operator and otherwise crossover is selected. (The procedure assumes that probability of inversion + probability of crossover = 1). Since the population consists of only five chromosomes, a candidate for applying inversion is selected by generating a random number between 1 and 5. Two random numbers are generated within the range of 'a' and 'b' and corresponding genes are selected for applying inversion. We find 'eval' of the new chromosome and if it is zero, then we have a solution. If the 'eval' is closer to zero than the highest 'eval' value of the chromosomes in the existing population of chromosomes, latter is replaced with the newly generated chromosome. If the 'eval' of the newly generated chromosome is not better than the existing 'eval', then we discard it. The 'eval' values are sorted and inversion is applied successively to get a sample of chromosomes of diverse nature.

Then, two chromosomes are selected randomly for applying crossover. The crossover point is selected as the location of the random number generated between one and the length of the chromosome. 'Eval' values of the new chromosomes are calculated and a comparison of these values with the 'eval' values of the existing chromosomes is done to check the feasibility of these chromosomes to be in the population. We accommodate either or both of these new chromosomes in the population, in place of the existing ones, depending upon the fitness of them. If the 'eval' values of either or both of these chromosomes are not better than the highest 'eval' value of the existing chromosomes, the new ones are discarded. Otherwise they are included in the population replacing the inferior ones. At the end of this procedure, the 'eval' values are sorted. As we have seen, after inversion or crossover, the selection of newly generated chromosomes to be in the next population depends on the superior fitness values of them. This was made possible by the use of a special type of selection procedure known as 'elitist selection'. Multiple chromosomes are found to have the same 'eval' value, which has the potential to lead to local optima or plateau, which we want to avoid. Here, we apply the concept of host parasite coevolution. All known repeated 'eval' values are treated as a separate family and labeled as reeval[]. These reeval[] values are considered as the attacks of parasites. When a new chromosome is evaluated and included in the population, the corresponding 'eval' value is stored in a separate family and labeled as eval[]. These eval[] values are taken as the defenses of the host. Thus, corresponding to the evolving attacks of parasite, we have the evolving defenses of host. This evolution of attacks of parasites and evolution of newer defenses by host has been implemented with the help of creating a special environment, which satisfies the condition

$$\prod_i (eval - reeval[i]) \neq 0 \text{ for } i=1, 2, 3,...... \quad (6)$$

Thus a new chromosome is evaluated for being in the population not just for having a better 'eval' value but the suitability to be fit in the newly created and evolving environment. When the chromosome passes these twin tests, we include the chromosome in the population and add the corresponding 'eval' value in the family eval[]. Thus, effectively we have an automatically defined function [Koza], which keeps on updating and acts as the evaluation function. In short, the fitness of a candidate is tested using equation (5) along with equation (6).

Thus, we have a composite evaluation function of an objective evaluation function and a subjective evaluation function [Watson & Pollack]. We call an evaluation function 'objective' if the fitness is depending only on that chromosome and is independent of other individuals. In the present work, the objective evaluation function corresponds to the evolutionary genetic algorithm, which checks the fitness of a chromosome using the equation (5). We call an evaluation function 'subjective' if the fitness of an individual is an outcome of its interactions with other individuals. In this work, the subjective evaluation function tests the suitability of the defenses of host which appear as a new eval[] value based on the equation (6). When a fresh attack of parasite is encountered in the form of unknown repeated reeval[] value, the methodology trains the host to be defensive to this attack. This is done by repeat inversion and a crossover to obtain new eval[] values. Thus, the attacks of parasites as reeval[] and defenses of host as eval[] evolve over the evolutionary process of searching through the fitness landscape. Hence, we have one population each of defenses of host and attacks of parasites, which reciprocally induce evolution. The pressure of this evolving 'arms race' is that the continued adaptations in some parasites as a new reeval[] value forces competitive adaptations in the defenses of the host to generate individuals with new eval[] value of increased performance. Effectively, the whole process works as a powerful optimization tool resulting in finding solution eventually.

**4.0 Implementation:**
The methods discussed in the paper have been implemented in C language. The chromosomes, the eval[] family and the reeval[] family are represented as arrays. The user supplies the initial conditions like probability of inversion, probability of crossover, range of solutions expected etc. The user also specifies the expected generation in which a solution could be reached. Within the expected generation, a number of evaluation functions are defined automatically, a number of attacks by parasites are encountered by the host and a number of defensive strategies are developed by the hosts.

**5.0 Experimental results:**
The result of running the program shows some important relationship between the number of attacks by parasites and the number of generations produced before the first solution is found. For demonstration purpose, we consider only a particular type of Diophantine equations, given by $x_1^n + x_2^n + ...... + x_n^n = N$, for n= 2, 3, ....., 10. The experimental set up uses the same initial conditions, which are given below, throughout the experiment.

    Probability of inversion : 0.60
    Probability of crossover: 0.40
    Range of solutions expected: between 1 and 25.

Initial population consists of five chromosomes, which depends on the type of equation we use.

| Diophantine Equation | Number of variables used | No of generations produced before the first solution | Number of attacks by parasites |
|---|---|---|---|
| $x_1^2 + x_2^2 = 149$ | 2 | 31 | 13 |
| $x_1^2 + x_2^2 + x_3^2 = 210$ | 3 | 57 | 22 |
| $x_1^2 + x_2^2 + .. + x_4^2 = 248$ | 4 | 49 | 19 |
| $x_1^2 + x_2^2 + .... + x_5^2 = 325$ | 5 | 50 | 19 |
| $x_1^2 + x_2^2 + .... + x_6^2 = 420$ | 6 | 139 | 60 |
| $x_1^2 + x_2^2 + .... + x_7^2 = 450$ | 7 | 132 | 56 |
| $x_1^2 + x_2^2 + .... + x_8^2 = 590$ | 8 | 93 | 40 |
| $x_1^2 + x_2^2 + .... + x_9^2 = 720$ | 9 | 52 | 23 |
| $x_1^2 + x_2^2 + .... + x_{10}^2 = 956$ | 10 | 319 | 147 |

Table 1: Results for equations with different number of variables

Table 1 gives result for different equations of varying number of variables and table 2 demonstrates the results for different equations of varying degrees. These results validate that irrespective of the varying number of variables and varying degrees, the methodology could give solutions of the Diophantine equations. They also show that the attacks by parasites are evolved in the evolution as the number of generations increase.

| Diophantine Equation | Degree of equation | No of generations produced before the first solution. | Number of attacks by parasites. |
|---|---|---|---|
| $x_1^2 + x_2^2 = 625$ | 2 | 24 | 9 |
| $x_1^3 + x_2^3 = 1008$ | 3 | 13 | 5 |
| $x_1^4 + x_2^4 = 706$ | 4 | 17 | 6 |
| $x_1^5 + x_2^5 = 1056$ | 5 | 24 | 9 |
| $x_1^6 + x_2^6 = 4097$ | 6 | 33 | 13 |
| $x_1^7 + x_2^7 = 2315$ | 7 | 1 | 0 |

Table 2: Results for equations with different degrees

Table 3 shows the values of attacks by parasites for the given sample of equations. These results demonstrate the diversity in attacks. The values of attacks fluctuate randomly. We cannot predict the value of the next attack in advance. These properties match completely with the nature of parasites' attack in nature.

| Sr. No | Equation | Value of attacks |
|---|---|---|
| 1 | $x_1^2 + x_2^2 = 149$ | 4, 36, 21, 53, 64, 101, 128, 32, 19, 75. |
| 2 | $x_1^2 + x_2^2 + x_3^2 = 210$ | 40, 16, 44, 4, 36, 19, 5, 28, 7, 6. |
| 3 | $x_1^2 + x_2^2 + x_3^2 + x_4^2 = 248$ | 18, 46, 14, 38, 21, 61, 44, 20, 25, 45. |
| 4 | $x_1^2 + x_2^2 + .... + x_5^2 = 325$ | 70, 77, 97, 62, 69, 47, 103, 19, 12, 20. |
| 5 | $x_1^2 + x_2^2 + .... + x_6^2 = 420$ | 281, 4, 22, 10, 28, 13, 14, 1, 7, 167. |
| 6 | $x_1^2 + x_2^2 + .... + x_7^2 = 450$ | 247, 41, 29, 47, 12, 49, 177, 210, 338, 117. |
| 7 | $x_1^2 + x_2^2 + .... + x_8^2 = 590$ | 306, 18, 34, 51, 99, 142, 282, 66, 274, 334. |
| 8 | $x_1^2 + x_2^2 + .... + x_9^2 = 720$ | 336, 63, 191, 95, 59, 107, 11, 83, 49, 16. |
| 9 | $x_1^2 + x_2^2 + .... + x_{10}^2 = 956$ | 451, 36, 20, 25, 29, 58, 422, 71, 34, 105. |

Table 3: Random nature of attacks

## 6.0 Conclusions and future work:

The paper explores the concept of reciprocally induced coevolution as a methodology offered in a typical genetic algorithm to find solutions of Diophantine equations. Experimental results indicate the random and diverse nature of attacks, which are in consonance with the attacks of parasites in nature. Currently, the authors are trying to extend this method as a paradigm to find numerical solutions of Diophantine equations with very large number of variables and large values as powers.